\documentclass[conference]{IEEEtran}
\IEEEoverridecommandlockouts
\usepackage{cite}
\usepackage{amsmath,amssymb,amsfonts}
\usepackage{algorithm}
\usepackage{algorithmic}
\usepackage{graphicx}
\usepackage{textcomp}
\usepackage{amsmath}
\usepackage{svg}
\usepackage{xcolor}
\usepackage{comment}
\def\BibTeX{{\rm B\kern-.05em{\sc i\kern-.025em b}\kern-.08em
  T\kern-.1667em\lower.7ex\hbox{E}\kern-.125emX}}

\begin{document}

\title{Performance and Cost Evaluation of Smart Contracts in Collaborative Health Care Environments*\\
\thanks{* This paper was achieved in cooperation with HP Brasil using incentives of Brazilian Informatics Law (Law n 8.248 of 1991). This study was financed in part by the Coordena\c{c}\~ao de Aperfei\c{c}oamento de Pessoal de N\'ivel Superior - Brasil (CAPES) - Finance Code 001. Avelino F. Zorzo is supported by CNPq (315192/2018-6). This work was supported by the INCT Forensic Sciences through the Conselho Nacional de Desenvolvimento Cient\'ifico e Tecnol\'ogico (CNPq $-$ process \# 465450/2014-8). Also, we thank to IFRS, DB Server and TokenHealth. }
}


\author{\IEEEauthorblockN{Roben Castagna Lunardi\IEEEauthorrefmark{1}\IEEEauthorrefmark{2},
Henry Cabral Nunes\IEEEauthorrefmark{1}, Vinicius da Silva Branco\IEEEauthorrefmark{1}, Bruno Hugentobler Lippert\IEEEauthorrefmark{1}, \\Charles Varlei Neu\IEEEauthorrefmark{1}\IEEEauthorrefmark{3}
Avelino Francisco Zorzo\IEEEauthorrefmark{1}}
\IEEEauthorblockA{\IEEEauthorrefmark{1}PUCRS, \IEEEauthorrefmark{2}IFRS,\IEEEauthorrefmark{3}UNISC - Brazil}
\IEEEauthorblockA{E-mail: \{roben.lunardi, henry.nunes, vinicius.branco, bruno.lippert@acad.pucrs.br, charles.neu\}@acad.pucrs.br, \\ avelino.zorzo@pucrs.br}}

\maketitle

\begin{abstract}
Blockchain emerged as a solution for data integrity, non-repudiation, and availability in different applications. Data sensitive scenarios, such as Health Care, can also benefit from these blockchain properties. Consequently, different research proposed the adoption of blockchain in Health Care applications. However, few are discussed about incentive methods to attract new users, as well as to motivate the system or application usage by existing end-users. Also, little is discussed about performance during code execution in blockchains. In order to tackle these issues, this work presents the preliminary evaluation of TokenHealth, an application for collaborative health practice monitoring with gamification and token-based incentives. The proposed solution is implemented through smart contracts using Solidity in the Ethereum blockchain. We evaluated the performance of both in Ropsten test network and in a Private instance. The preliminary results show that the execution of smart contracts takes less than a minute for a full cycle of different smart contracts. Also, we present a discussion about costs for using a Private instance and the public Ethereum main network.
\end{abstract}


\section{INTRODUCTION}\label{sec:intro}


Blockchain emerged as a promising technology after the proposition of the \textit{Bitcoin}~\cite{Nakamoto2008} cryptocurrency.
Additionally, blockchain has also been applied on solutions to solve problems on several other scenarios, such as Domain Name System (DNS) services ~\cite{Chang:2016}, storing and running programming code parts ~\cite{Ethereum2}, transaction control \cite{Min:2016}, electronic voting ~\cite{Moura:2017} and copyright control ~\cite{Kishigami:2015}. Many of these different applications have some requirements that are fulfilled by the adoption of blockchain, such as resilience (due to the decentralized characteristic of the network), non-repudiation (by using digital signatures in transactions) and tamper-resistance. Thus, blockchain provides reliability for the data it maintains.


Especially in the context of health data, there still are some problems regarding their handling, such as data that are often not recorded properly, out of date records, and even data may not be accessible by their owners, \textit{i.e.}, the end-users (patients)~\cite{Azaria:2016}. Thus, some studies suggest the use of blockchains to provide data integrity and availability \cite{Mettler:2016}\cite{Guo:2018}.
Although blockchain-based systems are used in different research that proposes health care monitoring solutions for digitized data sharing~\cite{Xia:2017}, storage~\cite{Mertz:2018} and access control~\cite{Rifi:2017}, there is little discussion about solutions that motivate the adoption and usage by end users. Such solutions could be based on techniques as bonuses\cite{CHEN-bonus-2018} or gamification \cite{Parizi:2018}, for example, and implemented using blockchains due to their ability on executing code in a distributed manner through smart contracts~\cite{Parizi:2018}~\cite{Zyskind:2015}. 

This paper aims to present and evaluate the TokenHealth project, a collaborative health practice monitoring system based on blockchain and smart contacts. Thus, the proposed system provides security through data integrity, resilience and availability. It also implements methods that motivate end-user adoption and usage. To evaluate performance in different environments, we used both the test network (Ropsten) and a private instance of Ethereum~\cite{Ethereum2}, which is currently one of the most popular blockchains that implement smart contracts. We also evaluate financial costs associated to those environments.

The remaining of this paper is structured as follows. Section \ref{sec:background} presents some background. Section \ref{sec:relatedwork} discusses some related work. Sections \ref{sec:casestudy} and \ref{sec:implementacao} present a case study, the proposed solution, describing its operation, details about the implementation, technologies and how smart contracts are used. Section \ref{sec:evaluation} presents and discusses the preliminary results. Finally, Section \ref{sec:final} concludes this paper and indicates some future work.

\section{Background} \label{sec:background}


With the popularization and success of Bitcoin, other blockchains emerged with different proposals and new technologies. Ethereum, like other blockchains that have a cryptocurrency (such as Bitcoin and Litecoin) associated, uses Proof-of-Work (PoW) as a consensus algorithm \cite{Tschorsch2016}. The consensus algorithm is a mechanism used to ensure that data addition follow a pre-combined business logic. This is necessary because blockchain works in a decentralized peer-to-peer (P2P) network where the nodes are unreliable and may act maliciously. Thus, the consensus algorithm \cite{roben-mobiquitous} ensures reliability to the data generated by any node in this untrusted environment.

Blockchain can have different characteristics depending on the purpose for the kind of application it was designed for. Zorzo \textit{et al.}~\cite{Zorzo:2018} proposed a layer-based model that show different solutions for communication, consensus algorithms, data management, and application layers. For example, hierarchical P2P architecture is better suited for IoT environments. One important feature regarding the application layer is the capability to support smart contracts, \textit{i.e.}, the capability to integrate the business logic of the application into the blockchain.

The concept of smart contracts was introduced in 1994 by Nick Szabo as a \textit{script} representing a contract that can enforce its terms automatically, reducing the need for intermediaries in the event of legal disputes \cite{Christidis:2016}.
Smart contracts are processed in a blockchain, being decentralized and allowing different models. It provides flexibility to process any application, providing immutability of the generated data, transparency on the operation and auditability on performed processes and transactions.


In Ethereum, each node has a virtual machine, called Ethereum Virtual Machine (EVM), which can be used to process \textit{bytecodes} representing smart contracts. Users can make special requests to the network by calling these smart contracts, allowing them to change their state or request information about the current state. Nodes process these requests based on the smart contract bytecode in its EVM and store the resulting smart contract state in the blockchain. This whole process is the same as the process of adding standard transactions and needs to be mined and verified by the whole network \cite{Ethereum2}.


\section{Related Work} \label{sec:relatedwork}

Different research proposed methods for rewarding, solving performance issues and adoption of smart contracts in blockchains. For example, Rouhani \textit{et al.}\cite{Rouhani:2019} discuss security issues and performance of smart contract execution based on different talks that intend to measure smart contracts performance. For example, some metrics are cited such as number of transactions per second, contract execution time, and block state update time. For measurement purposes, in this work, the average time to perform each contract will be adopted.

An approach to solve the mining work concentration problem by using a virtual currency service that proposes a new end-user usage incentive based on gamification instead of traditional economic incentives is presented by \cite{kano-gamif}. Experiments that show the feasibility of adopting the alternative incentive were presented with some discussion, regarding to the positive impact of psychological factors provided by gamification methods.

Parizi \textit{et al.}\cite{Parizi:2018} also discuss the gamification process in blockchain. They argue that gamification has been a trending topic to address human-centric concerns, specially in the online world, both in the industry and business and also in academic works. In this work, the authors also identified and discussed main human-related problems in decentralized blockchain systems and proposed a preliminarily gamified model, which is illustrated in the context of a typical blockchain system. 

Another work that discusses bonuses on blockchain-based systems is presented by Chen \textit{et al.} \cite{CHEN-bonus-2018}. Their method is built as a new type of decentralized bonus points alliance based on key technologies of blockchain, such as consensus mechanism and smart contract in blockchain and take the ``alliance blockchain''. This proposal takes advantage of technical features of decentralization, trust–consensus, distributed network, collective maintenance and advanced research on BonusPoints Alliance business model based on blockchain. This model is applied to design a system that could be used as a solution for the shortcomings of traditional alliance, such as a high cost of system  development, difficulty  of  bonus  points  exchange and difficulty of bonus points circulation.

A next generation repudiation system based on blockchain is proposed by Dennis \textit{et al.} \cite{repudiation-dennis}. The authors first discuss current reputation systems, current security vulnerabilities and how new blockchain-based technologies are currently used. Their goal is to propose a new reputation system based on blockchain technologies to solve problems that are, according to the authors, not yet solved on current generation reputation systems. Results are presented and discussed based on simulations. Performance is evaluated and limitations of the proposed solution are indicated and explained. Finally, the authors also present suggestions to overcome current limitations and indicate future directions.

\section{Case Study: TokenHealth} \label{sec:casestudy}

TokenHealth\footnote{Copyright and usage of the explicit business model presented in this paper is owned and reserved to TokenHealth: Diego Pirolla, Reider Arnaud Bernucio and Sergio Spacov.} is a system that aims to promote health through a collaborative tool, using methods based on \textit{tokens} and gamification (reputation system) functions. A proof of concept of a vaccination flow is implemented to validate and to evaluate the project. This proof of concept is intended to cover the entire vaccination cycle, from vaccine application, reapplication reminder, gamification and incentives.


\begin{figure*}[h!]
\centering\includegraphics[width=1\textwidth]{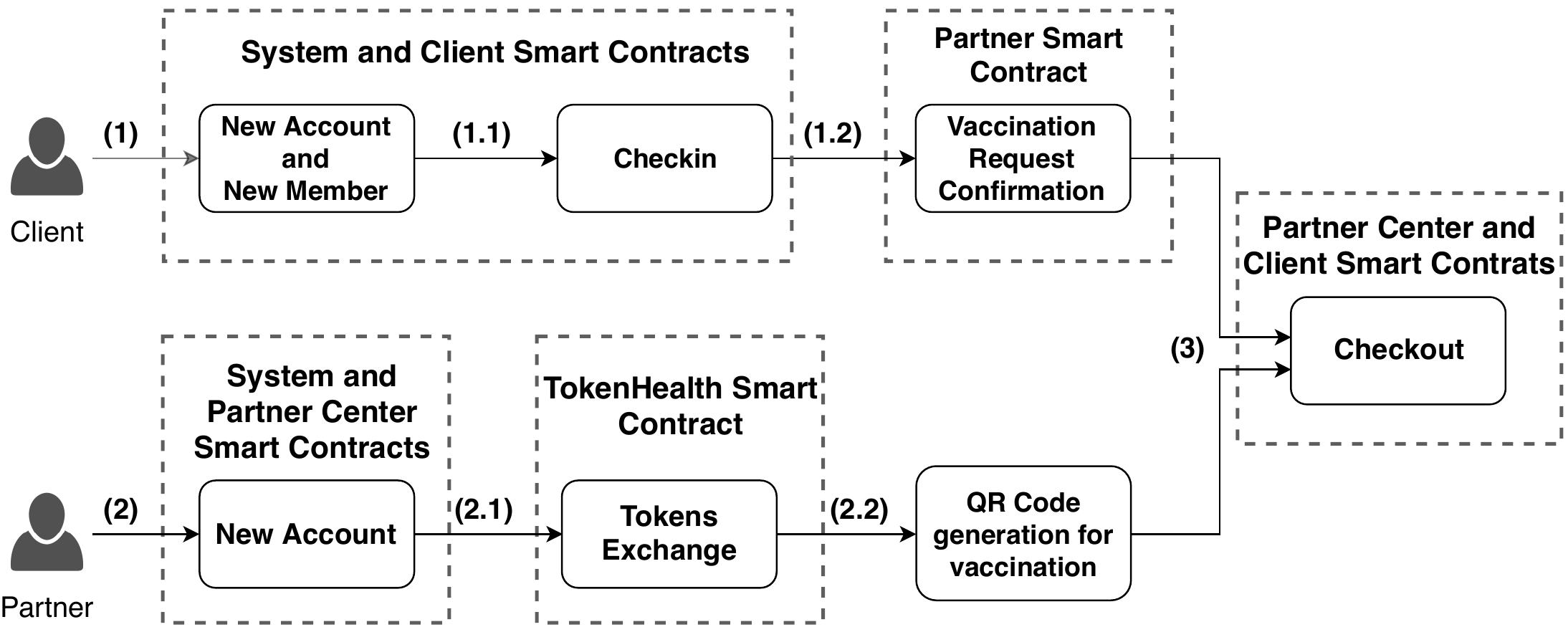}
\caption{Main flows for the vaccination system}
\label{fig:FiguraFluxo}
\end{figure*}

Also, this proof of concept implements security, by providing integrity, availability and transparency through the use of a blockchain. Additionally, \textit{Solidity}, a smart contract programming language, was used to implement the business rules in the blockchain. Thus, a generic model of operation is proposed, where users can keep their vaccines records updated and receive bonus for taking care of their health. Fig. \ref{fig:FiguraFluxo} presents an overview of the operation, flow of the main system components and the interaction with the actors that are involved.


The TokenHealth flow depends on two main actors: (\textit {(i)} the client, or their dependent, wants to be vaccinated and \textit {(ii)} a vaccination place (for example, a pharmacy, as illustrated in Fig. \ref{fig:FiguraFluxo}). The purpose of this flow is to connect clients and vaccination companies, encouraging the client to keep their vaccination grid up to date, and also to be rewarded with \textit{tokens}, which can be used to receive discounts on other purchases, for example. Thus, TokenHealth can also link customers to companies, improving fidelity. To do so, the blockchain technology plays a key role, as it allows the creation of transactions that persist data according to the business rules, as well as sending \textit{tokens}, a form of ``cryptocurrency'' that allows the exchange of values.


The flow starts with the customer and company registration (Flow 1 and 2, respectively, in Fig. \ref{fig:FiguraFluxo}). Through smart contracts, the customer then informs their personal data, registers their dependents and the vaccines that have already been received by each one. In another stream, the pharmacy registers its business data in ``System and Partner'' smart contract, and also informs its available vaccines that can be purchased. After that, the pharmacy can perform the purchase of \textit{tokens} that will be transferred to customers as bonus, as illustrated on Flow 2.1 in Fig. \ref{fig:FiguraFluxo}. TokenHealth provides many functionalities to the client, such as to list what vaccines should be taken, to select a partner (\textit{e.g.}, a pharmacy) that has the vaccines available, and to perform \textit{checkin} at the partner's place to inform that a vaccine has been taken. Also, the customer can choose to pay the full amount for the vaccine and receive \textit{tokens} or use their \textit {tokens} to get a discount, as shown in Flow 1.1 in Fig. \ref{fig:FiguraFluxo}

After the \textit{checkin} process is completed, the customer can go to the pharmacy, where a store attendant can see the customers \textit{checkin} on the ``TokenHealth Partner'' system, the partner side of TokenHealth. Then, an unique QR Code is generated for the current flow, so that the customer can read and confirm the release of the vaccine in TokenHealth. Flows 2.2 and 1.2, in Fig. \ref{fig:FiguraFluxo}, show this process. Once the vaccine is applied, the pharmacy can also confirm this process in ``TokenHealth Partner'', by using the\textit{checkout} process. Flow 3 illustrates this step.
The whole \textit{checkout} process is completed when both customer and pharmacy inform that the vaccine process is finished. If the customer has chosen to pay the full amount of the vaccine (without spending their \textit{tokens}), the partner must confirm it, so that the system then automatically sends rewarding \textit{tokens} to the customer, performing the bonus process.
On the other hand, if the customer has chosen to spend available \textit{tokens} as a payment method (and/or receive a discount, for example), after the partner confirmation, the customer must read a (new) QR Code to perform the \textit{checkout}. Thus, TokenHealth sends the \textit{tokens} to the partner automatically, in order to complete the payment.

\section{Implementation and technologies} \label{sec:implementacao}

\begin{figure}[h!]
\centering
\includegraphics[width=0.485\textwidth]{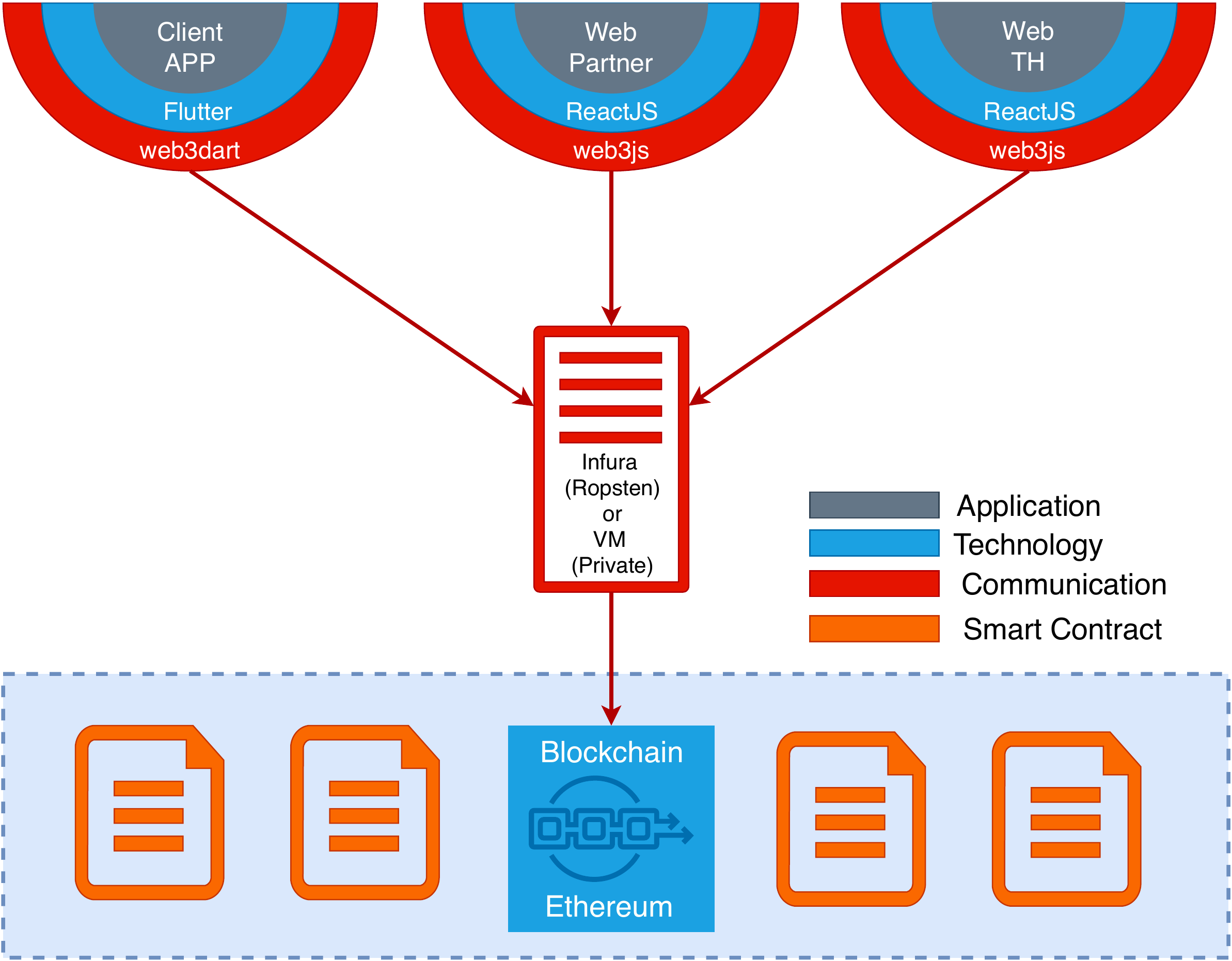}
\caption{Implementation in layers}
\label{fig:POC}
\end{figure}

To develop the proposed solution, a set of technologies was chosen. The development can be divided into four layers, as presented in Fig.~\ref{fig:POC}: (\textit{i}) developed applications, such as Client App and TH Web application (gray in the Fig.~\ref{fig:POC}); (\textit{ii}) technologies used for the application implementation, such as programming languages, libraries and Application Programming Interface (API) (in blue); (\textit{iii}) technologies used to communicate different technologies (in red); and (\textit{iv}) the smart contracts that implement the back end solution.

The \textit{web} interface was implemented using JavaScript as a programming language. This language was chosen because it is one of most used to connect through WEB3 (Ethereum interface) and has integration with the chosen libraries. For the client application, we used the \textit{Flutter} framework, as it generates applications for both \textit {Android} and \textit{iOS} platforms based on the same source code. Additionally, the Solidity language was chosen for the development of smart contracts, which is the main language on Ethereum. Also, the \textit{React.js} libraries were used to create the \textit{web} interface and \textit{Web3.js} to establish connection to the blockchain.

\section{Evaluation}\label{sec:evaluation}

In order to evaluate the performance of the proposed solution, we used the \textit{Ropsten Test Net}, a testing blockchain environment maintained by Ethereum. This environment is available to perform tests in an emulated environment, which contains similar characteristics of the main Public Ethereum network. An important advantage on evaluating the solution in Ropsten is that no financial investments are required, as Ropsten provides a faucet to request Ethers to this testing network. Additionally, we used a private instance of Ethereum in a cloud-based environment (using Google Cloud Platform~\cite{GoogleCloud} virtual machines). It is important to notice that we did not performed evaluation through the main Public Ethereum network due to financial costs associated with it.

One important advantage in private instances is that the mining difficulty can be set in the genesis block. This capability can help to start the blockchain with a difficulty that have a Proof-of-Work (PoW) adjusted to the infrastructure that will be used to maintain the blockchain. It is important because it allows to set a difficulty that enables time evaluation to produce new blocks in a higher throughput than in the main Public Ethereum network. 
 
We present a qualitative discussion in Table \ref{TabelaDiferencas} about different Ethereum options that can be used by the proposed solution. In our evaluation, both Ropsten and Private instances had a mining time lower than 1 minute. Consequently, the behaviour in both was similar. However, in the main Public Ethereum network, mining time is higher than 1 minute. It happens because of the high difficulty present in main Public Ethereum due to the dynamic difficulty increase over time, especially due to high computing power of the miners. However, when a distributed Application (dApp) uses a private instance of Ethereum, there is an associated infrastructure cost that should be considered. For example, for a small application, 5 nodes can be used to validate produced blocks. However, for larger applications, more nodes should be used to guarantee resilience and performance in the smart contracts execution.


\begin{table}[h!]
\centering
\caption{Ethereum networks comparison}
\begin{tabular}{|l|c|c|c|}
\hline
\multicolumn{1}{|c|}{}                            & \textbf{\begin{tabular}[c]{@{}c@{}}Ethereum \\ (Main)\end{tabular}} & \multicolumn{1}{l|}{\textbf{\begin{tabular}[c]{@{}l@{}}Ropsten \\(testnet)\end{tabular}}} & \multicolumn{1}{l|}{\textbf{Private Instance}} \\ \hline
\textbf{Mining Time}                         & \textgreater 5 minutes                          & \textless 1 minute                                        & \textless 1 minute               \\ \hline
\textbf{\begin{tabular}[c]{@{}l@{}}Mining Difficulty \\ \end{tabular}} & High                                   & Medium                                               & Settable                \\ \hline
\textbf{Financial Cost}                          & Yes (Ethers)                               & No                                                & Yes (infrastructure)              \\ \hline

\end{tabular}

\label{TabelaDiferencas}
\end{table}

Table~\ref{TabelaCustos} shows execution costs for each smart contract to present an overview of maintenance costs of the proposed solution. First, by analyzing the main Public Ethereum network, we estimated the cost in \textit{gas} (smart contract execution fee). The highest cost for an individual smart contract (as shown in Table ~\ref{TabelaCustos}) is the cost to create a new member, corresponding to a total of 0.002718 Ethers (or US\$0.453906, using the average quotation of \$167 dollars per Ether on September 26, 2019 \cite{CoinMarket}). A full user costs at least 0.003136 Ethers (sum of new account and one new member costs). Although this function has the highest cost, it should only occur once per user. 

The total cost for complete execution of the full vaccination cycle (checkin, confirmation and checkout) requires 0.000776 Ethers (or approximately \$0.129592). It is justified by the size and few processing required by the smart contracts used in the vaccination cycle. Those values are explained in Table \ref{TabelaCustos}.

\begin{table}[h!]
\centering
\caption{Smart Contracts execution costs}
\begin{tabular}{|l|c|c|}
\hline
\multicolumn{1}{|c|}{}   & \multicolumn{1}{l|}{\begin{tabular}{@{}c@{}}\textbf{Ethereum}\\ \textbf{(Main Network)}\end{tabular}} & \multicolumn{1}{l|}{\textbf{Private Instance}} \\ \hline
\textbf{\begin{tabular}{@{}c@{}}Create new account\end{tabular}} & \begin{tabular}{@{}c@{}}0.000418 Ethers\\ ($\sim$US\$0.069806)\end{tabular}         & -                        \\\hline
\textbf{Add a member}    & \begin{tabular}{@{}c@{}}0.002718 Ethers\\ ($\sim$US\$0.453906)\end{tabular}                          &  -                  \\ \hline
\textbf{Checkin}    & \begin{tabular}{@{}c@{}}0.000739 Ethers\\ ($\sim$US\$0.123413)\end{tabular}                         & -                   \\ \hline
\textbf{\begin{tabular}{@{}c@{}}Vaccination confirmation \end{tabular}}    & \begin{tabular}{@{}c@{}}0.000003 Ethers\\ ($\sim$US\$0.000501)\end{tabular}                          & -                   \\ \hline
\textbf{Checkout}    & \begin{tabular}{@{}c@{}}0.000034 Ethers\\ ($\sim$US\$0.005678)\end{tabular}                         & -                  \\ \hline
\textbf{\begin{tabular}{@{}c@{}}Full vaccination cycle\end{tabular}}  & \begin{tabular}{@{}c@{}}0.000776 Ethers\\ ($\sim$US\$0.129592)\end{tabular}         & -                        \\ \hline

\textbf{\begin{tabular}{@{}c@{}}Infrastructure (Monthly)\end{tabular}}    & -                         & US\$123,75                   \\ \hline
\end{tabular}
\label{TabelaCustos}
\end{table}

Another strategy to deploy a dApp is to use a private instance of Ethereum to maintain and execute smart contracts. For example, one can instantiate using cloud services with predefined infrastructure costs. For example, when allocating 5 specific machines to run Ethereum nodes in Google Cloud Platform ~\cite{GoogleCloud}, the monthly fixed cost would be around \$123.75 dollars (using 5 instances of \$24.75). It is worth noting that this cost represents only the basic configuration, \textit{i.e.}, it was not considered any elastic service or any kind of costs with maintenance and system configuration.

Some observations can be made when comparing the costs from both the main Public Ethereum network and a Private Instance. For example, Public Ethereum does not require maintenance cost and the cost for each execution is based on the number of cycles. However, using a private instance can allow a higher number of transactions with a fixed cost. As a comparison, the main Public Ethereum network can perform almost 1,000 full vaccination cycles for the same \$123.75 dollars (considering 26 September 2019 Ether exchange value~\cite{CoinMarket}). Considering the purpose of the developed application, one thousand cycles are not enough. Consequently, the main Public Ethereum network has a higher cost to perform the smart contracts considering a system only for vaccination. This discussion should be exploited in a future evaluation, considering other entities of health care, such as hospitals, health insurance, gym and others. 

\begin{figure*}[h!]
\centering
\includegraphics[width=1\textwidth]{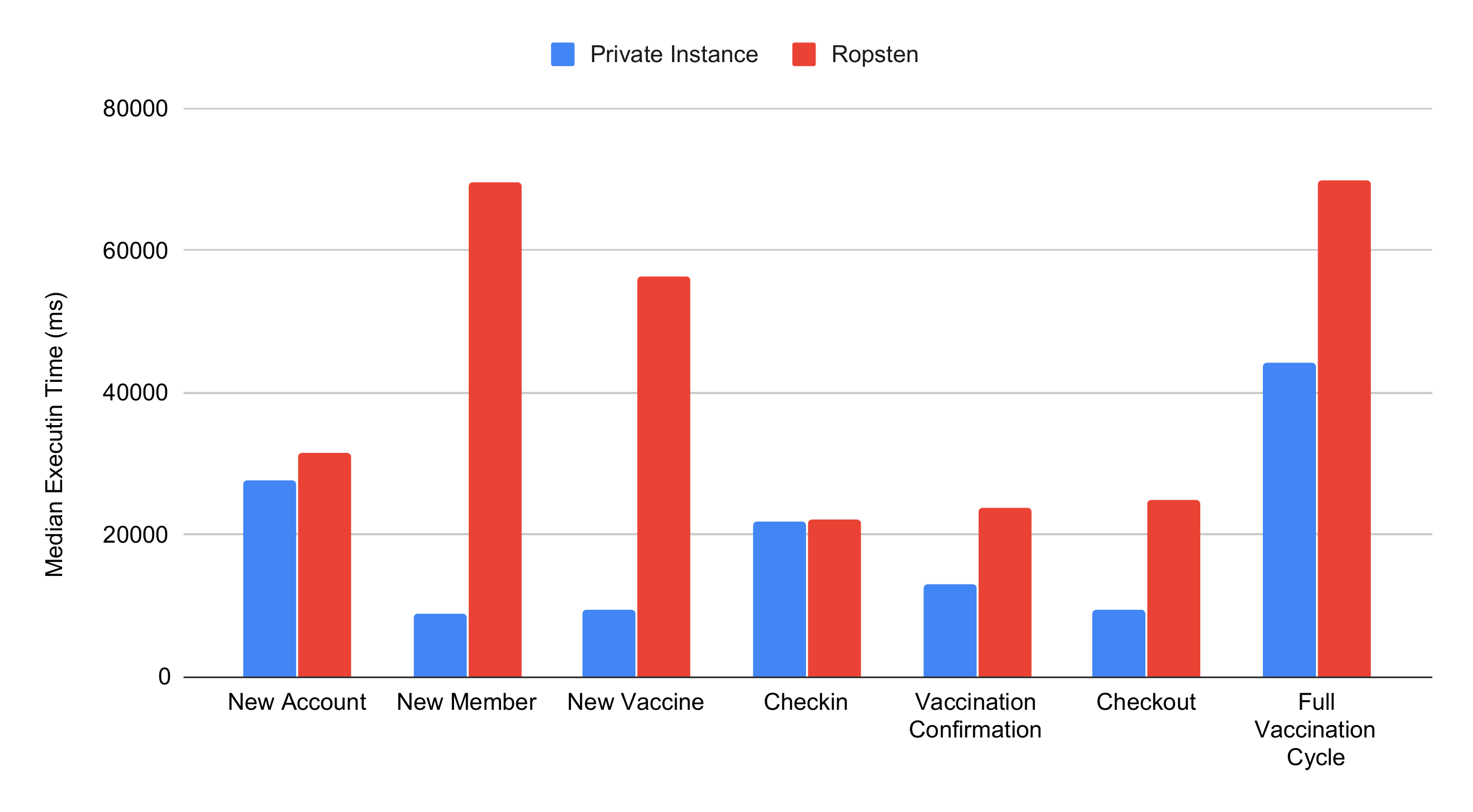}
\caption{Smart Contracts performance in Ropsten and Private Instance}
\label{fig:performance}
\end{figure*}

For the preliminary performance evaluation of the developed smart contracts, we used the Ropsten testing network and a Private Instance using Google Cloud~\cite{GoogleCloud} with 2 processing cores, 8GB of memory and 80GB of storage. The experiments were repeated 10 times and the results of the median executions are presented. Both in the private instance and in the Ropsten test network, good response time results related to the performance were obtained. An overview of executions can be observed in Fig.~\ref{fig:performance}.

We can observe, in Fig.~\ref{fig:performance}, that the execution of some smart contracts have a similar performance. For example, the smart contract to create a New Account executed in 27,744.5 milliseconds (median execution time) in a private instance and in 31,560.5 milliseconds in Ropsten, \textit{i.e.}, a difference of around 13\%. However, if we consider the execution of a smart contract with few processing requirements (shorter bytecode), the difference is higher. For example, the smart contract to add a New Member was executed in 8,906 milliseconds in the private instance and in 69,533.5 milliseconds in Ropsten. Also, it is important to note that Ropsten has a similar behavior present in the main Public Ethereum network, \textit{i.e.}, in some moments the throughput can be affected by problems such as fork resolutions or other issues. For the Full Vaccination Cycle, \textit{i.e.}, the sum of time spent in execution of Checkin, Vaccination Confirmation and Checkout smart contracts, it took 44,273.5ms in the private instance and 69,899.5ms in the Ropsten test network. The results demonstrates the viability, considering performance, to execute the main smart contracts for a dApp for vaccination. However, it was not possible to compare with the main Public Ethereum network due to the financial costs for acquiring Ethers.




\section{Final Considerations} \label{sec:final}


Health care is a recurring and important topic to the society, as several advances, new techniques, activities and medicines are constantly emerging. A plethora of systems and applications for health care and health activities monitoring are also currently available. However, it is important to create methods to promote end-users adoption and usage, in special for a collaborative approach that can help preventing diseases and health problems. Thus, this paper presents a solution for collaborative health economics systems using blockchain, exemplifying the usability of this technology in order to improve health and disease prevention through gamification and loyalty.

In addition, we presented some benefits of using blockchain in private instances or in blockchain public networks. As shown, the financial costs on the main Ethereum Public network are higher when the number of transactions is also high. However, when choosing to use private instance, the cost of infrastructure and personnel must be considered. Also, it was observed that the performance in the testing network were very similar to the values on the private instance, but no results were obtained with the main Ethereum Public network.
Finally, we can concluded that blockchain can be used as an alternative to a collaborative health monitoring system, as it makes the system safe by providing data immutability, ensuring that a business logic is preserved and the possibility of gamification by completing preventive health activities.

As a next step, we intend to expand the system to include medicines, medical consultations data and other activities regarding to preventive health. Also, we intend to expand the tests and to evaluate our solution on the main Ethereum Public network. Additionally, we intend to evaluate the Smart Contracts in different blockchains, especially in blockchains with different consensus algorithms, such as Hyperledger Fabric \cite{hyperledgerfabric:2016} and SpeedyChain~\cite{Lunardi:2018,Michelin:2018}. 

\bibliographystyle{IEEEtran}
\bibliography{icitst_sc}

\end{document}